\documentclass[letterpaper, 10 pt, conference]{ieeeconf}  

\IEEEoverridecommandlockouts                              

\usepackage{graphics} 
\usepackage{epsfig} 
\usepackage{times} 
\usepackage{amsmath} 
\usepackage{amssymb}  
\usepackage{mathbbol}
\usepackage{mathrsfs}
\usepackage{algorithm}
\usepackage{algorithmicx}
\usepackage{algpseudocode}
\usepackage{cite}
\usepackage{booktabs}
\usepackage{epstopdf}
\usepackage{subfigure,color,balance}
\usepackage{verbatim}
\usepackage{cases}
\usepackage{enumerate}
\usepackage{balance}
\usepackage{threeparttable}

\newcommand{\tr}{{{\mathsf T}}}

\newcommand{\method}[1]{\texttt{#1}}

\usepackage[colorlinks=true,      
linkcolor=black,      
citecolor=black,      
filecolor=black,      
urlcolor=blue]{hyperref}

\newtheorem{definition}{Definition}
\newtheorem{theorem}{Theorem}

\newtheorem{proposition}{Proposition}
\newtheorem{remark}{Remark}

\newtheorem{corollary}{Corollary}

\def\Y{{\bf Y}}

\def\0{{\bf 0}}
\def\1{{\bf 1}}

\def\col{{\mathrm {col}}}
\def\st{{\mathrm {subject~to}}}

\def\eg{{\em e.g.}}
\def\ie{{\em i.e.}}

\def\diag{{\rm diag}}

\title{\LARGE \bf
Data-Driven Predictive Control for Connected and Autonomous Vehicles in Mixed Traffic
}

\author{Jiawei Wang$^1$, Yang Zheng$^{2}$, Qing Xu$^{1}$ and Keqiang Li$^{1}$
\thanks{The work of J. Wang, Q. Xu and K. Li is supported by National Key R\&D Program of China with 2018YFE0204302, National Natural Science Foundation of China with 52072212, China Intelligent and Connected Vehicles (Beijing) Research Institute Co., Ltd., and Dongfeng Automobile Co., Ltd. All correspondence should be sent to Y.~Zheng and Q.~Xu.}
\thanks{$^{1}$J.~Wang, Q.~Xu and K.~Li are with the School of Vehicle and Mobility, Tsinghua University, Beijing, China. (wang-jw18@mails.tsinghua.edu.cn, \{qingxu,likq\}@tsinghua.edu.cn).}%
\thanks{$^{2}$Y. Zheng is with the Department of Electrical and Computer Engineering, University of California San Diego, CA 92093. ({zhengy@eng.ucsd.edu})}%
}

\begin{document}

\maketitle
\thispagestyle{empty}
\pagestyle{empty}

\begin{abstract}

Cooperative control of Connected and Autonomous Vehicles (CAVs) promises great benefits for mixed traffic. Most existing research focuses on model-based control strategies, assuming that car-following dynamics of  human-driven~vehicles are explicitly known. In this paper, instead of relying on a parametric car-following model, we introduce a data-driven predictive control strategy to achieve safe and optimal control for CAVs in mixed traffic. We first present a linearized dynamical model for mixed traffic systems, and investigate its controllability and observability. Based on these control-theoretic properties, we then propose a novel \method{DeeP-LCC} (Data-EnablEd Predictive Leading Cruise Control) strategy for CAVs based on measurable driving data to smooth mixed traffic. Our method is implemented in a receding horizon manner, in which input/output constraints are incorporated to achieve collision-free guarantees. Nonlinear traffic simulations reveal its saving of up to 24.96\% fuel consumption during a braking scenario of Extra-Urban Driving Cycle while ensuring safety.

\end{abstract}

\section{Introduction}

Connected and autonomous vehicles (CAVs) 
have provided new opportunities for smoothing traffic flow~\cite{guanetti2018control}. One typical technology is cooperative adaptive cruise control  that regulates a series of CAVs to achieve higher traffic efficiency and better fuel economy~\cite{li2017dynamical,zheng2016stability}. 
Given the gradual deployment of CAVs, there will be a transition phase of mixed traffic flow, where human-driven vehicles (HDVs) also exist. 
Recently, it has been shown theoretically and experimentally that incorporating HDVs' behavior into CAVs' controller design promises improved mixed traffic performance~\cite{zheng2020smoothing,stern2018dissipation,li2022cooperative,orosz2016connected}. 

The mixed traffic flow is essentially a complex human-in-the-loop cyber-physical system. 
Most existing research exploits microscopic car-following models to describe HDVs' behavior 
and designs model-based CAV control strategies~\cite{jin2017optimal,wang2020controllability,orosz2016connected}. 
In practice, however, human car-following dynamics are complex and nonlinear, which are non-trivial to identify accurately. Indeed, model-free or data-driven methods, 
bypassing model identifications, have recently received significant attention~\cite{recht2019tour,furieri2020learning}. For example, reinforcement learning~\cite{wu2021flow} and adaptive dynamic programming~\cite{huang2020learning} have been recently utilized for mixed traffic control, which employ large-scale driving data of HDVs to train control strategies of CAVs. However, these methods typically bring a heavy computation burden and have difficulties in including safety constraints in practical deployment. 

Recent advancements in data-driven predictive control have provided effective techniques towards safe learning~\cite{hewing2020learning}. One promising strategy is the recent Data-EnablEd Predictive Control (DeePC) method~\cite{coulson2019data}, which is able to achieve safe and optimal control for unknown systems using input/output measurements. Instead of identifying a parametric system model, DeePC relies on Willems' \emph{fundamental lemma}~\cite{willems2005note} to directly predict future trajectories. DeePC also
allows one to incorporate input/output constraints to ensure safety. 
Moreover, DeePC has shown its equivalence with sequential system identification and Model Predictive Control (MPC) for deterministic linear time-invariant (LTI) systems~\cite{coulson2019data}, and has demonstrated better control performance for nonlinear and non-deterministic systems~\cite{coulson2019regularized,dorfler2022bridging}. 
%
To our best knowledge, data-driven predictive control methods such as DeePC have not been utilized in mixed traffic control, and the results above are not directly applicable due to distinct dynamical properties of mixed traffic systems.

In this paper, we aim to design safe and optimal control strategies for CAVs to smooth mixed traffic flow that require no prior knowledge of HDVs' car-following dynamics.~In particular, motivated by DeePC~\cite{coulson2019data}, we introduce a Data-EnablEd Predictive Leading Cruise Control (\method{DeeP-LCC}) strategy. 
Our main contributions include: 1) We establish a linearized state-space model for a general mixed traffic system with multiple CAVs and HDVs under the Leading Cruise Control (LCC) framework~\cite{wang2021leading}. We define  measurable driving data as system output, highlighting the fact that HDVs' equilibrium spacing is practically unknown. This issue has been neglected in many results on mixed traffic that require state-feedback control~\cite{jin2017optimal,wang2020controllability,huang2020learning}. We show that the linearized mixed traffic system is not completely controllable unless the first vehicle is a CAV, but is observable under a mild condition. These control-theoretic results serve as the foundation for our reformulation of DeePC for mixed traffic.
2) We propose a \method{DeeP-LCC} method, in which the CAVs utilize measurable driving data for optimal and safe controller design without identifying an explicit parametric car-following model. The standard DeePC requires the underlying system to be controllable~\cite{coulson2019data,willems2005note}, and thus cannot be directly applied to mixed traffic. To address this issue, we introduce an external input signal to record the data of the head vehicle. Together with CAVs' control input, this contributes to full controllability. 
	Spacing constraints are also incorporated on the driving behavior for safety guarantees. 

The rest of this paper is organized as follows. Section~\ref{Sec:2} introduces the mixed traffic modeling, and Section~\ref{Sec:3} presents the controllability and observability results. We then present \method{DeeP-LCC} in Section~\ref{Sec:4} and  traffic simulations in Section~\ref{Sec:5}. This paper is concluded in Section~\ref{Sec:6}.


\section{Theoretical Modeling Framework}
\label{Sec:2}

\begin{figure}[t]
	\vspace{1mm}
	\centering
	\includegraphics[scale=0.32]{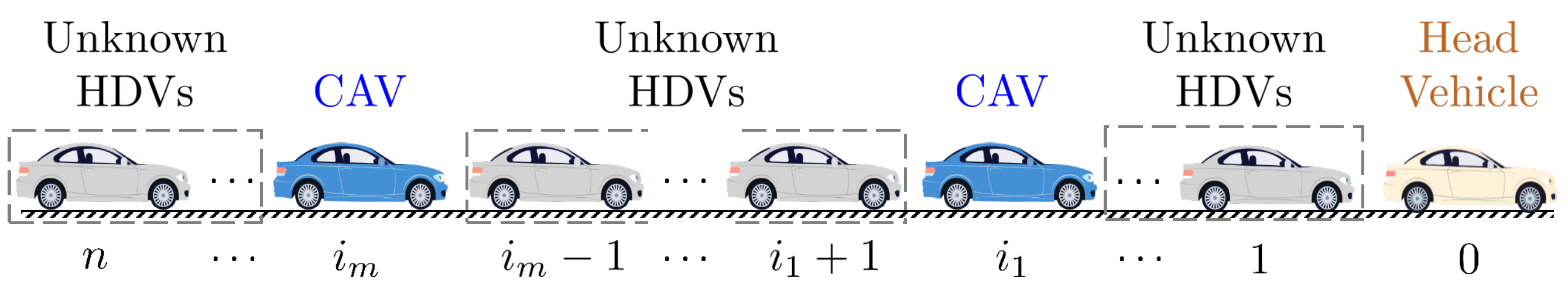}
	\vspace{-5mm}
	\caption{Schematic of mixed traffic flow. The head vehicle, at the very beginning, is indexed as 0, behind which there exist $n$ vehicles consisting of $m$ CAVs and $n-m$ HDVs with unknown driving dynamics.}
	\label{Fig:SystemSchematic}
	\vspace{-5mm}
\end{figure}

In this section, we first introduce the parametric modeling of HDVs' car-following behavior, and then present the linearized dynamics of a general mixed traffic system under the LCC framework; see~\cite{wang2021leading} for a detailed motivation of LCC. 

As shown in Fig.~\ref{Fig:SystemSchematic}, we consider a general mixed traffic system with $n+1$ individual vehicles, among which there exist one head vehicle, indexed as $0$, and $m$ CAVs and $n-m$ HDVs. 
Define $\Omega=\{1,2,\ldots,n\}$ as the set of all the vehicle indices, and $S=\{i_1,i_2,\ldots,i_m\}\subseteq \Omega$ as the set of the CAV indices, where $i_1 < i_2 < \ldots < i_m$ also represent the spatial locations of the CAVs in mixed traffic flow. The position, velocity and acceleration of the $i$-th vehicle at time $t$ is denoted as $p_i(t)$, $v_i(t)$ and $a_i(t)$, respectively. 

\subsection{Car-Following Dynamics of HDVs}
\label{Sec:CarFollowingModel}

There exist many well-established continuous-time models for car-following dynamics, including the optimal velocity model (OVM), the intelligent driver model (IDM) and their variants~\cite{orosz2010traffic}. 
Most of these models can be written in the following nonlinear form
\begin{equation}\label{Eq:HDVModel}
\dot{v}_i(t)=F\left(s_i(t),\dot{s}_i(t),v_i(t)\right), \quad i \in \Omega \backslash S,
\end{equation}
where $s_i(t)=p_{i-1}(t)-p_i(t)$ denotes the car-following spacing of vehicle $i$, 
and $\dot{s}_i(t)=v_{i-1}(t)-v_i(t)$ denotes the relative velocity. The nonlinear function $F(\cdot)$ represents that the acceleration of an HDV depends on the relative distance, relative velocity and its own velocity. 

In an equilibrium traffic state, each vehicle moves with the same equilibrium velocity $v^{*}$ and the corresponding equilibrium spacing $s^{*}$. According to~\eqref{Eq:HDVModel}, we have
\begin{equation} \label{Eq:Equilibrium}
F\left(s^*,0,v^*\right)=0. 
\end{equation}
%
%
Upon assuming that the mixed traffic flow is around an equilibrium state~$(s^*,v^*) $, we define the error state between actual and equilibrium point as
$\tilde{s}_i(t)=s_i(t)-s^*,\, \tilde{v}_i(t)=v_i(t)-v^*,$ $i \in \Omega$, 
where $\tilde{s}_{i}$, $\tilde{v}_{i}$ represent the spacing error and velocity error of vehicle $i$, respectively. Then a linearized dynamics model for each HDV can be derived by using~\eqref{Eq:Equilibrium} and applying the first-order Taylor expansion to~\eqref{Eq:HDVModel} 
\begin{equation}\label{Eq:LinearHDVModel}
\begin{cases}
\dot{\tilde{s}}_i(t)=\tilde{v}_{i-1}(t)-\tilde{v}_i(t),\\
\dot{\tilde{v}}_i(t)=\alpha_{1}\tilde{s}_i(t)-\alpha_{2}\tilde{v}_i(t)+\alpha_{3}\tilde{v}_{i-1}(t),\\
\end{cases} \quad i \notin S,
\end{equation}
where $\alpha_{1} = \frac{\partial F}{\partial s}, \alpha_{2} = \frac{\partial F}{\partial \dot{s}} - \frac{\partial F}{\partial v}, \alpha_{3} = \frac{\partial F}{\partial \dot{s}}$ with the partial derivatives evaluated at the equilibrium state ($s^*,v^*$). To reflect the real stable driving behavior of human drivers, we have $\alpha_{1}>0$, $\alpha_{2}>\alpha_{3}>0$. More linearization details can be found in \cite{jin2017optimal,zheng2020smoothing}.

\subsection{State-Space Model of Mixed Traffic Systems}
\label{Section:TrafficModel}

Similar to~\cite{zheng2020smoothing,jin2017optimal,huang2020learning}, we use the acceleration of each CAV as the control input, \ie,
$\dot{v}_i(t)=u_i(t), \, i \in S.$
%
A second-order model is used to describe the linear longitudinal dynamics of each CAV
\begin{equation} \label{Eq:LinearCAVModel}
	\begin{cases}
	\dot{\tilde{s}}_i(t)=\tilde{v}_{i-1}(t)-\tilde{v}_i(t),\\
	\dot{\tilde{v}}_i(t)=u_i(t),\\
	\end{cases} \quad i\in S.
\end{equation}

To derive a linearized model of the mixed traffic system shown in Fig.~\ref{Fig:SystemSchematic}, we lump the error states of all the vehicles as the mixed traffic system state ($x(t)\in \mathbb{R}^{2n}$)
\begin{equation*}\label{Eq:SystemState}
x(t)=\begin{bmatrix}\tilde{s}_{1}(t),\tilde{v}_{1}(t),\tilde{s}_{2}(t),\tilde{v}_{2}(t),\ldots,\tilde{s}_{n}(t),\tilde{v}_{n}(t)\end{bmatrix}^{\tr},
\end{equation*}
and lump the acceleration signal of all the CAVs as the collective control input
$ u(t) = \begin{bmatrix}u_{i_{1}}(t), u_{i_{2}}(t), \ldots, u_{i_{m}}(t)\end{bmatrix}^{\tr}$ $\in \mathbb{R}^m$. 
Then, based on the linearized HDVs' car-following model~\eqref{Eq:LinearHDVModel} and the CAV's dynamics~\eqref{Eq:LinearCAVModel}, the linearized state-space model for the mixed traffic system is obtained as
\begin{equation} \label{Eq:LinearSystemModel}
\dot{x}(t)=Ax(t)+Bu(t)+H\epsilon(t),
\end{equation}
where $\epsilon(t)=\tilde{v}_{0}(t)=v_0(t)-v^*$, \ie, the velocity error of the head vehicle, is regarded as an external input to the system. The system matrices in~\eqref{Eq:LinearSystemModel} are given by (see~\cite{wang2021leading} for details)
\begin{align*}
A&=\begin{bmatrix} A_{1,1} & & & &   \\
A_{2,2} & A_{2,1} & &  &   \\
& \ddots& \ddots&  &  \\
& & A_{n-1,2}& A_{n-1,1} &  \\
& & &  A_{n,2}&A_{n,1}\\
\end{bmatrix} \in \mathbb{R}^{2n\times 2n} ,\\
B &= \begin{bmatrix}
\mathbb{e}_{2n}^{2i_{1}}, \mathbb{e}_{2n}^{2i_{2}}, \ldots, \mathbb{e}_{2n}^{2i_{m}}
\end{bmatrix}\in \mathbb{R}^{2n \times m},\\
H &= \begin{bmatrix}
h_{1}^{\tr},h_{2}^{\tr},\ldots,h_n^{\tr}
\end{bmatrix}^{\tr}\in \mathbb{R}^{2n \times 1},
\end{align*}
where $\mathbb{e}_{p}^r$ denotes a $p \times 1$ unit vector, with the $r$-th entry being one and the others being zeros, and 
\begin{align*}
A_{i,1} &= \begin{cases}
P_1, \; i\notin S;\\
S_1, \; i\in S;
\end{cases} \;
A_{i,2} = \begin{cases}
P_2, \; i\notin S;\\
S_2, \; i\in S;
\end{cases}\\
h_{1} &= \begin{bmatrix} 1 \\ \alpha_{3}\end{bmatrix},\;
h_j= \begin{bmatrix} 0 \\ 0 \end{bmatrix},\; j \in \{2,3,\ldots,n\},
\end{align*}
with
\begin{align*}
\setlength{\arraycolsep}{3pt}
P_{1} \!=\! \begin{bmatrix} 0 & -1 \\ \alpha_{1} & -\alpha_{2} \end{bmatrix}\!,
P_{2} \!=\! \begin{bmatrix} 0 & 1 \\ 0 & \alpha_{3} \end{bmatrix}\!,
	S_1\! =\! \begin{bmatrix} 0 & -1 \\ 0 & 0 \end{bmatrix}\!,
S_2\! =\! \begin{bmatrix} 0 & 1 \\ 0 & 0 \end{bmatrix}\!.
\end{align*}


\noindent \textbf{Measurable driving data:} Note that the state in~\eqref{Eq:LinearSystemModel} cannot be directly measured. In practice, the equilibrium velocity $v^*$ can be obtained from the steady-state velocity of the head vehicle. 
However, the equilibrium spacing $s^*$ for the HDVs is non-trivial to accurately estimate, since the car-following behavior of each human driver is unknown. It is thus impractical to observe the information of the spacing error signal of the HDVs, \ie, $\tilde{s}_i$ $(i \notin S)$. For the CAVs, by contrast, their equilibrium spacing can be designed~\cite{zheng2020smoothing}, and thus their spacing error signal can be directly measured. Accordingly, we introduce an output signal $y(t)$ as follows
\begin{equation*} 
y(t)=\begin{bmatrix}\tilde{v}_{1}(t),\tilde{v}_{2}(t),\ldots,\tilde{v}_{n}(t),\tilde{s}_{i_1}(t),\tilde{s}_{i_2}(t),\ldots,\tilde{s}_{i_m}(t)\end{bmatrix}^{\tr},
\end{equation*}
where $y(t) \in \mathbb{R}^{n+m}$ consists of the velocity errors of both the HDVs and the CAVs, \ie, $\tilde{v}_i$ $(i \in \Omega)$, and the spacing errors of all the CAVs, \ie, $\tilde{s}_i$ $(i \in S)$. The output in~\eqref{Eq:LinearSystemModel} is given by
\begin{equation} \label{Eq:SystemOutput}
y(t)=Cx(t),
\end{equation}
where 
$
C=\begin{bmatrix}
\mathbb{e}_{2n}^{2}, \mathbb{e}_{2n}^{4}, \ldots, \mathbb{e}_{2n}^{2n},\mathbb{e}_{2n}^{2i_1-1},\mathbb{e}_{2n}^{2i_2-1},\ldots,\mathbb{e}_{2n}^{2i_m-1}
\end{bmatrix}^\tr$ $\in \mathbb{R}^{(n+m) \times 2n}
$ is the output matrix. 

\section{Controllability and Observability \\ of Mixed Traffic Systems}
\label{Sec:3}

Controllability and observability serve as foundations in dynamical systems~\cite{skogestad2007multivariable}. For mixed traffic systems, existing research has revealed the controllability for the scenario of one single CAV and multiple HDVs, \ie, $|S|=1$~\cite{wang2021leading,jin2017optimal}. These results have been unified in the recent LCC framework with one single CAV~\cite{wang2021leading}. 

For general mixed traffic systems with multiple HDVs and CAVs, given by~\eqref{Eq:LinearSystemModel} and~\eqref{Eq:SystemOutput}, 
we have the following results.

\begin{theorem}[Controllability]
	\label{Theorem:Controllability}
	Consider the general mixed traffic system  given by~\eqref{Eq:LinearSystemModel} and~\eqref{Eq:SystemOutput}, where there exist $m$ $(m \geq 1)$ CAVs with indices $S=\{i_1,i_2,\ldots,i_m\}$, $i_1 < i_2 < \ldots < i_m$. The following statements hold.
	\begin{enumerate}
		\item When $1 \in S$, \ie, $i_1 = 1$, the mixed traffic system is controllable if the following condition holds
		\begin{equation}
	\label{Eq:ControllabilityCondition}
	\alpha _{1}- \alpha _{2} \alpha _{3}+ \alpha _{3}^{2} \neq 0.
	\end{equation}
		\item When $1 \notin S$, the mixed traffic system is not completely controllable but is stabilizable,~if~\eqref{Eq:ControllabilityCondition} holds. The subsystem consisting of the states $\tilde{s}_{1},\tilde{v}_{1},\ldots,\tilde{s}_{i_1-1},\tilde{v}_{i_1-1}$ is not controllable but is stable, while the subsystem consisting of the states $\tilde{s}_{i_1},\tilde{v}_{i_1},\ldots,\tilde{s}_{n},\tilde{v}_{n}$ is controllable.
	\end{enumerate}
\end{theorem}

The proof idea is based on controllability invariance under state feedback with respect to the result in~\cite[Theorem 2]{wang2021leading}; the technical proof is presented in an extended version~\cite{wang2022deeplcc}. 
Theorem~\ref{Theorem:Controllability} indicates that the general mixed traffic system is not completely controllable unless the vehicle immediately behind the head vehicle is a CAV. This is expected, since the motion of the HDVs between the head vehicle and the first CAV cannot be influenced by the CAVs' control.

\begin{theorem}[Observability]
	\label{Theorem:Observability}
	The general mixed traffic system,  given by~\eqref{Eq:LinearSystemModel} and~\eqref{Eq:SystemOutput}, is observable when~\eqref{Eq:ControllabilityCondition} holds.
\end{theorem}

The observability result can be proved by adapting~\cite[Theorem 4]{wang2021leading}. The slight asymmetry between Theorems~\ref{Theorem:Controllability} and~\ref{Theorem:Observability} is due to the fact that the control input in \eqref{Eq:SystemOutput} includes only the acceleration signal of all the CAVs, while the system output consists of the velocity error signal of all the vehicles and the spacing error signal of the CAVs. Theorem~\ref{Theorem:Observability} reveals the observability of the full state $x(t)$ in mixed traffic under a mild condition. This observability result facilitates the design of our \method{DeeP-LCC} controller. 

Still, the controllability of a dynamical system is a desired property, which guarantees the data-driven behavior representation in the next section. 
Note that the velocity error signal of the head vehicle $ \epsilon(t)=v_0(t)-v^*$ is an external input. This signal is not controlled, but can be measured in practice. Define $
	\hat{u}(t) = \col \left(\epsilon (t),u (t) \right)
$
as a combined input signal and $\widehat{B} = \begin{bmatrix} H,B
\end{bmatrix}$ as the corresponding input matrix. Then, a reformulated model for the mixed traffic system is
\begin{equation} \label{Eq:TransformedMixedTrafficSystem}
	\begin{cases} 
\dot{x}(t)=Ax(t)+\widehat{B} \hat{u}(t),\\
y(t)=Cx(t).
\end{cases}
\end{equation}
For this system model, we have the following result.
\begin{corollary}
	\label{Corollary:TransformedSystemControllability}
	Consider the mixed traffic system given by~\eqref{Eq:TransformedMixedTrafficSystem}, where there exist $m$ $(m \geq 1)$ CAVs. Then, the system $(A, \widehat{B}, C)$ is controllable and observable if~\eqref{Eq:ControllabilityCondition} holds.
\end{corollary}


\section{\method{DeeP-LCC} for Mixed Traffic Flow}
\label{Sec:4}

In this section, we first introduce a non-parametric data-centric representation of the mixed traffic system behavior based on Willems' fundamental lemma~\cite{willems2005note}, and then present the Data-EnablEd Predicted Leading Cruise Control (\method{DeeP-LCC}) strategy for mixed traffic control.

\subsection{Non-Parametric Representation of System Behavior}

The system model in \eqref{Eq:LinearSystemModel} and~\eqref{Eq:SystemOutput} is  continuous. It can be straightforwardly  transformed into the discrete time domain 
\begin{equation} \label{Eq:DT_TrafficModel}
\begin{cases}
x(k+1) = A_\mathrm{d}x(k) + B_\mathrm{d}u(k) + H_\mathrm{d} \epsilon(k),\\
y(k) = C_\mathrm{d}x(k),
\end{cases}
\end{equation}
where $
A_\mathrm{d} = e^{A\Delta t}\in \mathbb{R}^{2n\times 2n},B_\mathrm{d}=\int_{0}^{\Delta t} e^{A t} Bd t\in \mathbb{R}^{2n\times m},H_\mathrm{d}=\int_{0}^{\Delta t} e^{A t} Hd t \in \mathbb{R}^{2n\times 1},C_\mathrm{d}=C \in \mathbb{R}^{(n+m)\times 2n} 
$, and $\Delta t>0$ is the sampling interval.

Model-based control strategies typically follow the sequential system identification and control procedure: learning the parametric model (\ie, $A_\mathrm{d},B_\mathrm{d},H_\mathrm{d},C_\mathrm{d}$ in~\eqref{Eq:DT_TrafficModel}) in advance and then performing model-based controller design. By contrast, the recently proposed DeePC method~\cite{coulson2019data} is a \textit{non-parametric method} that bypasses the system identification process and directly designs the control input compatible with historical system data. In particular, DeePC directly uses past data to predict the system ``behavior'' based on Willems' \emph{fundamental lemma}~\cite{willems2005note}.  
The following notion of persistent excitation is needed.
\begin{definition}
	The signal  $\omega = \col (\omega_1,\omega_2,\ldots,\omega_T)$ of length $T$ is persistently exciting of order $l$ $(l \leq T)$ if the following Hankel matrix is of full row rank
	\begin{equation}
		\mathcal{H}_{l}(\omega):=\begin{bmatrix}
			\omega_{1} &\omega_{2}& \cdots & \omega_{T-l+1} \\
			\omega_{2} &\omega_{3}& \cdots & \omega_{T-l+2}\\
			\vdots & \vdots & \ddots & \vdots \\
			\omega_{l} &\omega_{l+1}& \cdots & \omega_{T}
		\end{bmatrix}.
	\end{equation}
\end{definition}

\vspace{1mm}
\noindent\textbf{Data collection:} The standard DeePC requires the system to be completely controllable. Thus, we rely on the reformulated system model~\eqref{Eq:DT_TrafficModel} with two input signals $u,\epsilon$ combined together to design our \method{DeeP-LCC} method for mixed traffic. We begin by collecting a sequence trajectory data of length $T$ from the system~\eqref{Eq:DT_TrafficModel} with sampling interval $\Delta t$. The collected data includes: 1) the combined input sequence $\hat{u}^\mathrm{d}=\col (\hat{u}^\mathrm{d}(1),\ldots,$ $\hat{u}^\mathrm{d}(T))\in \mathbb{R}^{(m+1)T}$, consisting of CAVs' acceleration sequence $u^\mathrm{d}=\col (u^\mathrm{d}(1),\ldots,$ $u^\mathrm{d}(T))\in \mathbb{R}^{mT}$ and the velocity error sequence of the head vehicle $\epsilon ^\mathrm{d}=\col (\epsilon^\mathrm{d}(1),\ldots,\epsilon^\mathrm{d}(T)) \in \mathbb{R}^{T}$;
2) the output sequence of mixed traffic $y ^\mathrm{d}=\col (y^\mathrm{d}(1),\ldots,y^\mathrm{d}(T)) \in \mathbb{R}^{(n+m)T}$.

These data samples could be generated offline, or collected online from the trajectory data of those involved vehicles. Then, we partition the pre-collected data into two parts, corresponding to ``past data'' of length $T_{\mathrm{ini}} \in \mathbb{N}$ and ``future data'' of length $N \in \mathbb{N}$. Precisely, define 
\begin{equation}
\begin{gathered}
\label{Eq:DataHankel}
\begin{bmatrix}
U_{\mathrm{p}} \\
U_{\mathrm{f}}
\end{bmatrix}:=\mathcal{H}_{T_{\mathrm{ini}}+N}(u^{\mathrm{d}}), \quad \begin{bmatrix}
E_{\mathrm{p}} \\
E_{\mathrm{f}}
\end{bmatrix}:=\mathcal{H}_{T_{\mathrm{ini}}+N}(\epsilon^{\mathrm{d}}),\\
\begin{bmatrix}
Y_{\mathrm{p}} \\
Y_{\mathrm{f}}
\end{bmatrix}:=\mathcal{H}_{T_{\mathrm{ini}}+N}(y^{\mathrm{d}}),
\end{gathered}    
\end{equation}
where $U_{\mathrm{p}}$ and $U_{\mathrm{f}}$ consist of the first $T_{\mathrm{ini}}$ block rows and the last $N$ block rows of $\mathcal{H}_{T_{\mathrm{ini}}+N}(u^{\mathrm{d}})$, respectively (similarly for $E_{\mathrm{p}}, E_{\mathrm{f}}$ and $Y_{\mathrm{p}}, Y_{\mathrm{f}}$). 

\vspace{1mm}
\noindent\textbf{System behavior representation:} motivated by Willems' fundamental lemma~\cite{willems2005note} and the DeePC formulation~\cite{coulson2019data}, we have the following result: given time $t$, we define $u_{\mathrm{ini}}=\col(u(t-T_{\mathrm{ini}}),u(t-T_{\mathrm{ini}}+1),\ldots,u(t-1))$ as the control input sequence within a past time horizon of length $T_{\mathrm{ini}}$, and $u= \col(u(t),u(t+1),\ldots,u(t+N-1))$ as the control input sequence within a predictive time horizon of length $N$ (similarly for $\epsilon_\mathrm{ini},\epsilon$ and $y_\mathrm{ini},y$).

\begin{proposition}
\label{Proposition:DeePCMixedTraffic}
Suppose that~\eqref{Eq:ControllabilityCondition} holds, and the combined input sequence $\hat{u}^\mathrm{d}$ is persistently exciting of order $T_{\mathrm{ini}}+N+2n$. Then, any trajectory of the mixed traffic system~\eqref{Eq:DT_TrafficModel} of length $T_{\mathrm{ini}}+N$, denoted as $\col (u_\mathrm{ini},\epsilon_\mathrm{ini},y_\mathrm{ini},u,\epsilon,y)$, can be constructed via
\begin{equation}
\label{Eq:AdaptedDeePCAchievability}
\begin{bmatrix}
U_\mathrm{p} \\ E_\mathrm{p}\\Y_\mathrm{p} \\ U_\mathrm{f} \\ E_\mathrm{f}\\ Y_\mathrm{f}
\end{bmatrix}g=
\begin{bmatrix}
u_\mathrm{ini} \\ \epsilon_\mathrm{ini}\\ y_\mathrm{ini} \\ u \\\epsilon \\ y
\end{bmatrix},
\end{equation}
where $g\in \mathbb{R}^{T-T_\mathrm{ini}-N+1}$. If $T_{\mathrm{ini}} \geq 2n $, $y$ is uniquely determined from~\eqref{Eq:AdaptedDeePCAchievability}, $\forall (u_\mathrm{ini} ,\epsilon_\mathrm{ini}, y_\mathrm{ini},u,\epsilon)$. 
\end{proposition}

This proposition is adapted from Willems' fundamental lemma~\cite{willems2005note} and the DeePC method~\cite{coulson2019data} for the mixed traffic system~\eqref{Eq:DT_TrafficModel}. It  reveals that we can use past trajectories to predict the future trajectory of the mixed traffic system without identifying an explicit parametric model. 
Specifically, given a past trajectory $(u_\mathrm{ini},\epsilon_\mathrm{ini},y_\mathrm{ini})$ and a future input sequence $(u,\epsilon)$, the formulation~\eqref{Eq:AdaptedDeePCAchievability} allows one to predict the future output sequence $y$ directly from pre-collected data $(u^\mathrm{d},\epsilon^\mathrm{d},y^\mathrm{d})$. Therefore, we can bypass a parametric system model and directly use non-parametric data-centric representation for the behaviors of the mixed traffic system.

Note that the velocity error of the head vehicle $\epsilon (t)$ is under human control. It is always oscillating around zero in practice considering the real behavior of human drivers, \ie, the drivers always attempt to maintain the equilibrium velocity while also suffering from small perturbations. Accordingly, given a trajectory with length
$
     T \geq (m+1)(T_{\mathrm{ini}}+N+2n)-1
$,
and persistently exciting acceleration input signal $u(t)$ of CAVs (\eg, white noise with zero mean), the persistent excitation requirement in Proposition~\ref{Proposition:DeePCMixedTraffic} for the combined input $\hat{u}(t)$ is naturally satisfied.

\begin{remark}
Willems' fundamental lemma~\cite{willems2005note} reveals that the subspace consisting of all valid trajectories is identical to the range space of the data Hankel matrix of the same order generated by sufficiently rich inputs. DeePC has recently applied this fundamental lemma to predictive control~\cite{coulson2019data}. However, DeePC requires the underlying system to be controllable. 
For mixed traffic control, we introduce the external input signal $\epsilon$, \ie, the velocity error of the head vehicle, to make the reformulated system controllable, and rely on~\eqref{Eq:AdaptedDeePCAchievability} for representation of the mixed traffic system behavior. In addition, for an observable system, one can estimate the system state from past data whose length is not smaller than the state dimension. Given the observability of the mixed traffic system, the underlying initial state for the future trajectory is implicitly fixed from~\eqref{Eq:AdaptedDeePCAchievability} when $T_{\mathrm{ini}} \geq 2n $, which guarantees the uniqueness property in Proposition~\ref{Proposition:DeePCMixedTraffic}; see~\cite{coulson2019data} for more discussions on DeePC.
\end{remark}



\subsection{Design of Cost Function and Constraints in \method{DeeP-LCC}}

Here, we show how to utilize the non-parametric behavior representation~\eqref{Eq:AdaptedDeePCAchievability} to design the control input of the CAVs in mixed traffic flow, motivated by the standard DeePC~\cite{coulson2019data}. Precisely, we aim to design the future behavior $(u,\epsilon,y)$ for the mixed traffic system through a receding horizon manner based on pre-collected data $(u^\mathrm{d},\epsilon^\mathrm{d},y^\mathrm{d})$ and the most recent past data $(u_\mathrm{ini},\epsilon_\mathrm{ini},y_\mathrm{ini})$ which are updated online. 

The past external input sequence $\epsilon_\mathrm{ini}$ can be collected in the control process, but the future external input sequence $\epsilon$ cannot be designed (it is controlled by a human driver). Although the future human behavior might be predicted using ahead traffic conditions, it is still non-trivial to achieve accurate prediction. Considering that the driver always attempts to maintain the equilibrium velocity, 
one natural approach is to assume that the future velocity error of the head vehicle is zero, \ie,
\begin{equation} \label{Eq:FutureExternalInput}
\epsilon = \mathbb{0}_N,
\end{equation}
where $\mathbb{0}_N$ denotes an $N \times 1$ vector of all zeros.

Similar to the LCC framework~\cite{wang2021leading}, we consider the performance of the entire mixed traffic system for CAVs' controller design. Precisely, we consider a quadratic-form cost function $J(y,u)$, which penalizes the output deviation from the equilibrium state and the energy of control input from time $t$, defined as follows
\begin{equation} \label{Eq:CostDefinition}
J(y,u) = \sum\limits_{k=t}^{t+N-1}\left( \left\|y(k)\right\|_{Q}^{2}+\left\|u(k)\right\|_{R}^{2}\right), 
\end{equation}
where the coefficient matrices $Q,R$ are set as 
$
Q=\diag(Q_v,Q_s)
$
with $Q_v = \diag(w_v,\ldots,w_v) \in \mathbb{R}^{n\times n}$, $Q_s = \diag(w_s,\ldots,w_s) \in \mathbb{R}^{m\times m}$ and $R = \diag(w_u,\ldots,w_u) \in \mathbb{R}^{m\times m} $, where $w_v,w_s,w_u$ represent the penalty for the velocity errors of all the vehicles, spacing errors of all the CAVs, and control inputs of the CAVs, respectively.

\begin{algorithm}[t]
	\caption{\method{DeeP-LCC}}
	\label{Alg:DeeP-LCC}
	\begin{algorithmic}[1]
		\Require
		Pre-collected traffic data $(u^{\mathrm{d}},\epsilon^{\mathrm{d}},y^{\mathrm{d}})$, initial time $t_0$, terminal time $t_f$;
		\State Construct data Hankel matrices $U_\mathrm{p} , U_\mathrm{f}, E_\mathrm{p} , E_\mathrm{f}, Y_\mathrm{p} , Y_\mathrm{f}$;
		\State Initialize past traffic data $(u_{\mathrm{ini}},\epsilon_{\mathrm{ini}},y_{\mathrm{ini}})$ before the initial time $t_0$;
		\While{$t_0 \leq t \leq t_f$}
		\State Solve~\eqref{Eq:AdaptedDeePCforNonlinearSystem} for optimal predicted input $u^*=\col(u^*(t),u^*(t+1),\ldots,u^*(t+N-1))$;
		\State Apply the input $u(t) \leftarrow u^*(t)$ to the CAVs;
		\State $t \leftarrow t+1$ and update past traffic data $(u_{\mathrm{ini}},\epsilon_{\mathrm{ini}},y_{\mathrm{ini}})$;
		\EndWhile
	\end{algorithmic}
\end{algorithm}

We further incorporate several constraints for safety guarantees. In particular, a minimal spacing constraint for CAVs is required to avoid rear-end collisions. Accordingly, we impose a lower bound $\tilde{s}_\mathrm{min}$ on the spacing error:
$
	\tilde{s}_{i} \geq \tilde{s}_\mathrm{min}, \, i\in S
$. 
In addition, we note that existing CAV controllers tend to leave an extremely large spacing from the preceding vehicle in the control procedure (see, \eg,~\cite{stern2018dissipation} and the discussions in~\cite[Section V-D]{wang2020controllability}), which in practice might cause vehicles from adjacent lanes to cut in. To address this problem, we also include an upper bound: $
	\tilde{s}_{i} \leq \tilde{s}_\mathrm{max}, \, i\in S.
$
The imposed bounds on the spacing errors are then converted to the following compact form on the system output $y$
\begin{equation} \label{Eq:SafetyConstraint}
	\tilde{s}_\mathrm{min} \leq I_{(n+m)N} \otimes \begin{bmatrix}
	\mathbb{0}_{m \times n} & I_m
	\end{bmatrix} y \leq \tilde{s}_\mathrm{max}.
\end{equation}
%
In addition, the control input of each CAV is constrained as 
\begin{equation} \label{Eq:AccelerationConstraint}
		a_\mathrm{min} \leq u \leq a_\mathrm{max},
\end{equation} 
where $	a_\mathrm{min}$ and $a_\mathrm{max}$ denote the minimum and the maximum acceleration, respectively.

\subsection{Formulation of \method{DeeP-LCC}}

Motivated by DeePC~\cite{coulson2019data}, we are now ready to present the following optimization problem to obtain the optimal control input of the CAVs at each time step
\begin{equation} \label{Eq:AdaptedDeePC}
\begin{aligned}
\min_{g,u,y} \quad &J(y,u) \\
\st \quad &\eqref{Eq:AdaptedDeePCAchievability},\eqref{Eq:FutureExternalInput},\eqref{Eq:SafetyConstraint},\eqref{Eq:AccelerationConstraint}.
\end{aligned}
\end{equation}
One significant distinction of~\eqref{Eq:AdaptedDeePC} from the standard DeePC is the introduction of the future velocity error sequence $\epsilon$ of the head vehicle, \ie, the external input of the mixed traffic system. Note that $\epsilon$ is not a decision variable in~\eqref{Eq:AdaptedDeePC}, unlike $u$ and $y$. Instead, it is fixed as constant, as shown in~\eqref{Eq:FutureExternalInput}. 

For implementation, the optimal control problem~\eqref{Eq:AdaptedDeePC} is solved in a receding horizon manner, similarly to standard MPC. 
Unlike MPC, by contrast, the optimal control problem~\eqref{Eq:AdaptedDeePC} does not rely on an explicit parametric system model, but utilizes the data-centric representation~\eqref{Eq:AdaptedDeePCAchievability} to predict future behaviors. However, the formulation~\eqref{Eq:AdaptedDeePCAchievability} is only valid for deterministic LTI mixed traffic systems. In practice, the car-following behavior of HDVs is nonlinear (see Section~\ref{Sec:CarFollowingModel}), and also has uncertainties, leading to a nonlinear and non-deterministic system. Moreover, practical traffic data is always noise-corrupted. Hence, the equality constraint~\eqref{Eq:AdaptedDeePCAchievability} becomes inconsistent for practical traffic flow. 

Similar to the regulation for standard DeePC~\cite{coulson2019data}, we introduce a slack variable $\sigma_y \in \mathbb{R}^{(n+m)T_\mathrm{ini}}$ for the system past output to ensure the feasibility of the equality constraint, yielding the following optimization problem
  \begin{equation} \label{Eq:AdaptedDeePCforNonlinearSystem}
 \begin{aligned}
 \min_{g,u,y,\sigma_y} \quad &J(y,u)+\lambda_g \left\|g\right\|_2^2+\lambda_y \left\|\sigma_y\right\|_2^2\\
 \st \quad & \begin{bmatrix}
 U_\mathrm{p} \\ E_\mathrm{p}\\Y_\mathrm{p} \\ U_\mathrm{f} \\ E_\mathrm{f}\\ Y_\mathrm{f}
 \end{bmatrix}g=
 \begin{bmatrix}
 u_\mathrm{ini} \\ \epsilon_\mathrm{ini}\\ y_\mathrm{ini} \\ u \\\epsilon \\ y
 \end{bmatrix}+\begin{bmatrix}
 0\\0\\ \sigma_y \\0 \\0 \\0
 \end{bmatrix},\\ &\eqref{Eq:FutureExternalInput},\eqref{Eq:SafetyConstraint},\eqref{Eq:AccelerationConstraint},
 \end{aligned}
 \end{equation}
 which is suitable for nonlinear and non-deterministic mixed traffic flow. This is our final \method{DeeP-LCC} formulation at each time step. 
In~\eqref{Eq:AdaptedDeePCforNonlinearSystem}, the slack variable $\sigma_y$ is penalized with a weighted two-norm penalty function, and the weight coefficient $\lambda_y>0$ can be chosen sufficiently large such that $\sigma_y \neq 0$ only if the equality constraint is infeasible. In addition, a two-norm penalty on $g$ with a weight coefficient $\lambda_g>0$ is also incorporated to avoid overfitting in case of noise-corrupted data samples. 
As discussed in~\cite{huang2021decentralized,coulson2019regularized}, such regulation on $g$ coincides with distributional two-norm robustness. The receding horizon implementation of \method{DeeP-LCC} is listed in Algorithm~\ref{Alg:DeeP-LCC}.

\begin{remark}
In our \method{DeeP-LCC} for mixed traffic, we introduce the external input signal and utilize~\eqref{Eq:FutureExternalInput} to straightforwardly predict its future value. 
Besides~\eqref{Eq:FutureExternalInput}, another potential approach to address the unknown future external input is to assume a bounded future velocity error of the head vehicle. This idea is similar to robust DeePC against unknown external disturbances; see, \eg,~\cite{huang2021decentralized,lian2021adaptive}. 
It is an interesting topic for future research to design robust \method{DeeP-LCC} for mixed traffic when the head vehicle is oscillating around a particular equilibrium velocity.
\end{remark}

\section{Traffic Simulations}
\label{Sec:5}

We now present nonlinear and non-deterministic traffic simulations to validate the performance of \method{DeeP-LCC}. Our simulation setup is motivated from the standard Extra-Urban Driving Cycle (EUDC)~\cite{dieselnet2013emission}. The nonlinear OVM model in~\cite{wang2020controllability} is used for HDVs~\eqref{Eq:HDVModel}, and a noise signal following the uniform distribution of $\mathbb{U}[-0.1,0.1]$ is added to the OVM model\footnote{Our code is available at \url{https://github.com/soc-ucsd/DeeP-LCC}.}.

\subsection{Experimental Setup}

We consider eight vehicles with two CAVs, \ie, $n=8$, $m=2$ in Fig.~\ref{Fig:SystemSchematic}. The two CAVs are located at the third and the sixth vehicles respectively, \ie, $S=\{3,6\}$. 
In \method{DeeP-LCC}, we use the following parameters. 
1) In offline data collection: the length for the pre-collected data is $T=2000$ with $\Delta t = 0.05 \,\mathrm{s}$. We collect these data around  $15\,\mathrm{m/s}$, and there exists a uniformly distributed signal of $\mathbb{U}[-1,1]$ on both $u^\mathrm{d}$ and  $\epsilon^\mathrm{d}$. This naturally satisfies the persistent excitation requirement in Proposition~\ref{Proposition:DeePCMixedTraffic}. 2) In
online predictive control: the time horizons for the future and past trajectories are set to $N=50$, $T_{\mathrm{ini}}=20$, respectively. For constraints, we have $\tilde{s}_{\max} = 20, \tilde{s}_{\min} = -15,a_{\max} = 2, a_{\min} = -5$. In the cost function~\eqref{Eq:CostDefinition}, the weight coefficients are set to $w_v=1,w_s=0.5,w_u=0.1$, and the regularized parameters in~\eqref{Eq:AdaptedDeePCforNonlinearSystem} are set to $\lambda_g=100,\lambda_y=10000$.

For the HDVs' OVM model, we assume a heterogeneous parameter setup around the nominal value. We also consider the standard output-feedback MPC for comparison in our simulations, assuming that the explicit parametric linearized model is known.  The corresponding parameter setup remains the same as that in \method{DeeP-LCC}. 
Note that the traffic flow could have different equilibrium states in different time periods. We use the average velocity of the head vehicle among the past horizon of $T_{\mathrm{ini}}$ 
as the estimated equilibrium velocity, and the corresponding equilibrium spacing is manually set; see the extended version~\cite{wang2022deeplcc} for more details.

\subsection{Numerical Results}
\label{Sec:Simulation3}

\begin{figure}[t]
	\vspace{1mm}
	\centering
	\subfigure[All HDVs]
	{
	\includegraphics[scale=0.42]{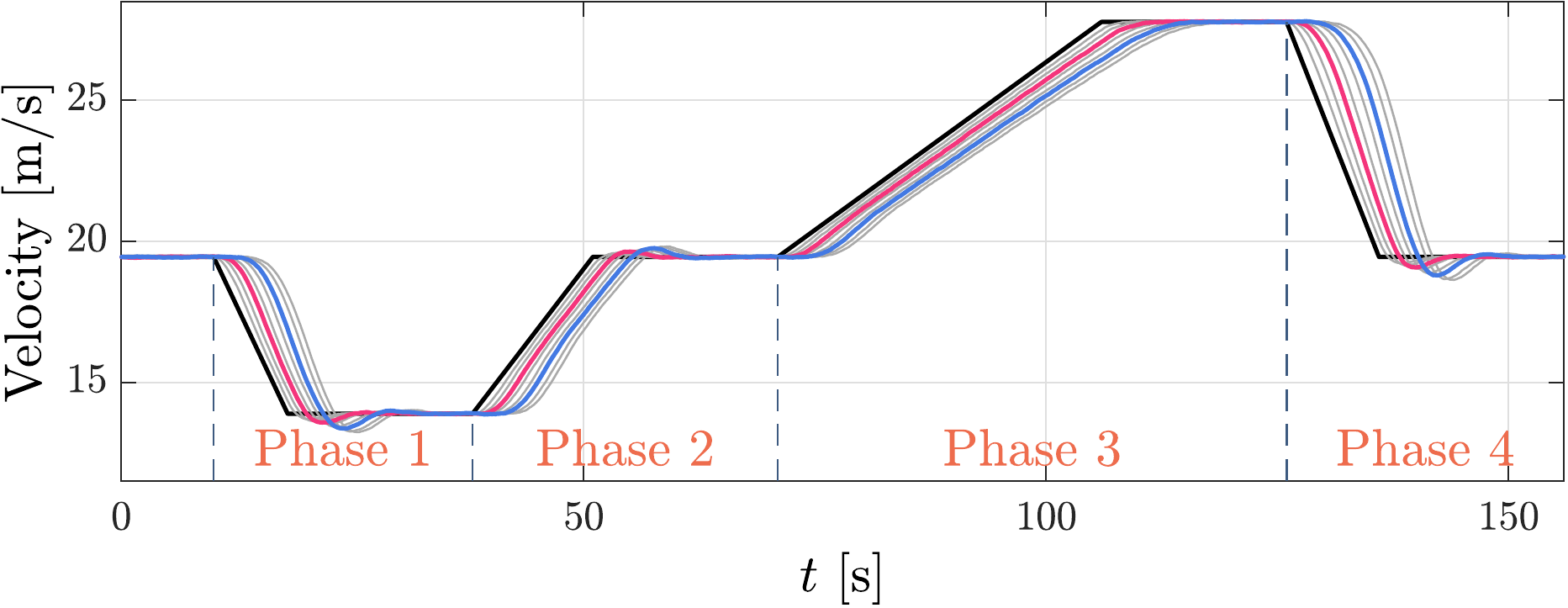}
	}
	\subfigure[\method{DeeP-LCC}]
	{
	\includegraphics[scale=0.42]{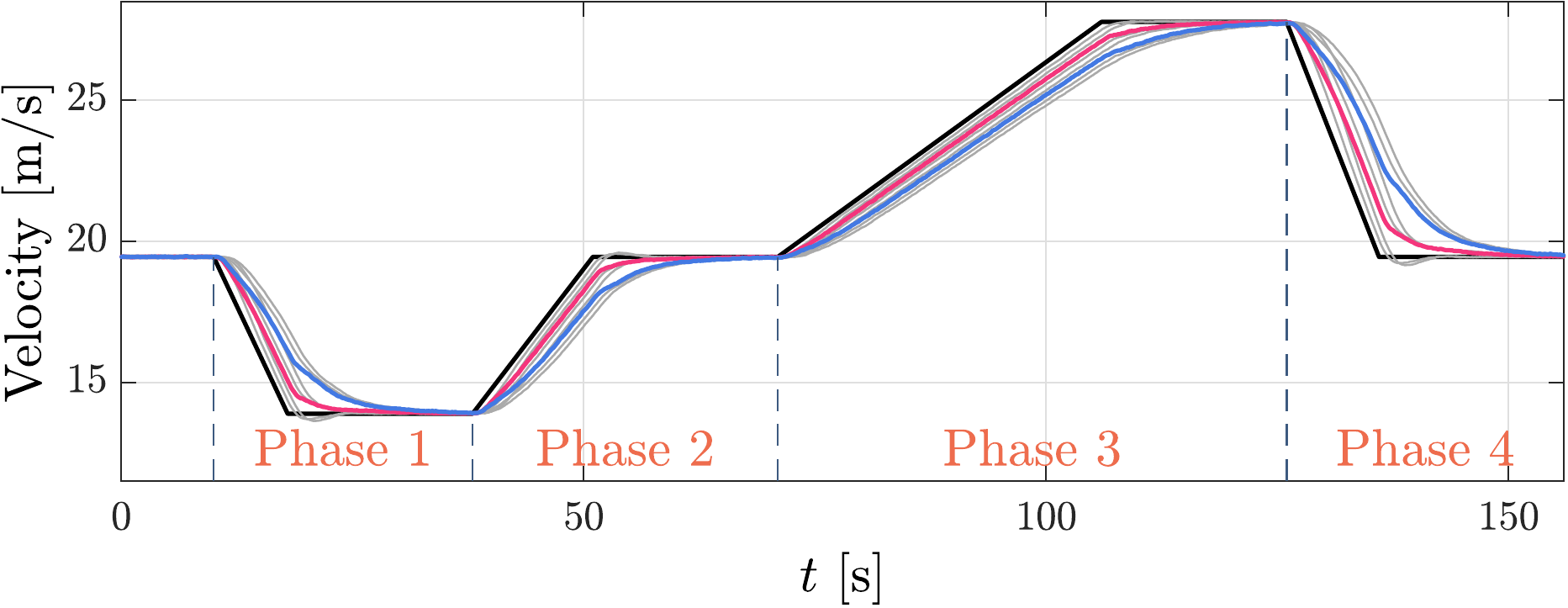}
	}
	\vspace{-2mm}
	\caption{Velocity profiles in Experiment A. The black profile represents the head vehicle, and the gray profile represents the HDVs. The red profile and the blue profile represent the first and the second CAV, respectively. (a) All the vehicles are HDVs. (b) The CAVs utilize \method{DeeP-LCC}.}
	\label{Fig:NEDCSimulation}
	\vspace{-5mm}
\end{figure}

\noindent \textbf{Experiment A}: We first design a comprehensive simulation scenario to validate the capability of \method{DeeP-LCC} in improving traffic performance. Specifically, motivated by EUDC driving cycle, we design a velocity trajectory for the head vehicle (see the bleck profile in Fig.~\ref{Fig:NEDCSimulation}). For performance evaluation, we calculate the fuel consumption using the numerical model in~\cite{bowyer1985guide} for the vehicles indexed from 3 to 8, given that the first two HDVs would not be influenced.

\begin{table}[t]
	\begin{center}
		\caption{Fuel Consumption in Experiment A}\label{Tb:NEDC_FC}
		\begin{threeparttable}
		\setlength{\tabcolsep}{2mm}{
		\begin{tabular}{cccc}
		\toprule
			& All HDVs & MPC & \method{DeeP-LCC} \\\hline
			Phase 1 & 172.59 & 158.79 ($\downarrow$7.99\%) & 159.17 ($\downarrow$7.78\%) \\
			Phase 2 & 379.13 & 374.35 ($\downarrow$1.26\%) & 374.60 ($\downarrow$1.19\%) \\
			Phase 3 & 817.16 & 812.91 ($\downarrow$0.52\%) & 812.71 ($\downarrow$0.54\%) \\
			Phase 4 & 399.86 & 377.66 ($\downarrow$5.55\%) & 377.58 ($\downarrow$5.57\%) \\
			Total Simulation & 1977.16 & 1928.19 ($\downarrow$2.48\%) & 1929.09 ($\downarrow$2.43\%) \\
			\bottomrule
		\end{tabular}}
		\begin{tablenotes}
		\footnotesize
		\item[1] All the values have been rounded and the unit is $\mathrm{mL}$ in this table.
		\end{tablenotes}
		\end{threeparttable}
	\end{center}
	\vspace{-5mm}
\end{table}

The simulation results are shown in Fig.~\ref{Fig:NEDCSimulation}. It can be clearly observed that compared to the case where all the vehicles are HDVs, \method{DeeP-LCC} apparently mitigates velocity perturbations and smooths traffic flow with only two CAVs existing in mixed traffic. The results of the fuel consumption is presented in Table~\ref{Tb:NEDC_FC}, with the whole simulation separated into four phases (as clarified in Fig.~\ref{Fig:NEDCSimulation}). Both MPC and \method{DeeP-LCC} reduce the fuel consumption throughout the four phases, and particularly, the two controllers contribute to a greater improvement on traffic performance in the braking phases (Phases 1 and 4) than the accelerating phases (Phases 2 and 3). In particular, \method{DeeP-LCC} saved $7.78\%$ and $5.57\%$ fuel consumption during Phases 1 and 4, respectively.

Note that the MPC controller utilizes the nominal model to design the control input, while the \method{DeeP-LCC} controller relies on the trajectory data to directly predict the future system behavior. In practice, MPC might be  inapplicable, since the nominal model for individual HDVs is non-trivial to identify. By contrast, \method{DeeP-LCC} achieves similar performance compared to MPC using only trajectory data, without explicitly identifying a parametric model. Hence, \method{DeeP-LCC}  has demonstrated great potential to improve traffic performance in practical mixed traffic flow.

\vspace{1mm}

\noindent \textbf{Experiment B}: To further validate the safety performance of \method{DeeP-LCC}, we design a particular braking scenario motivated by EUDC, where the head vehicle takes a sudden emergency brake with maximum deceleration capability, maintains the low velocity for a while, and finally accelerates to the original normal velocity. This is a typical emergency case in real traffic flow, which requires the CAVs' control to have strict safety guarantees from rear-end collision.

\begin{figure}[t]
	\vspace{1mm}
	\centering
	\subfigure[]
	{\includegraphics[scale=0.42]{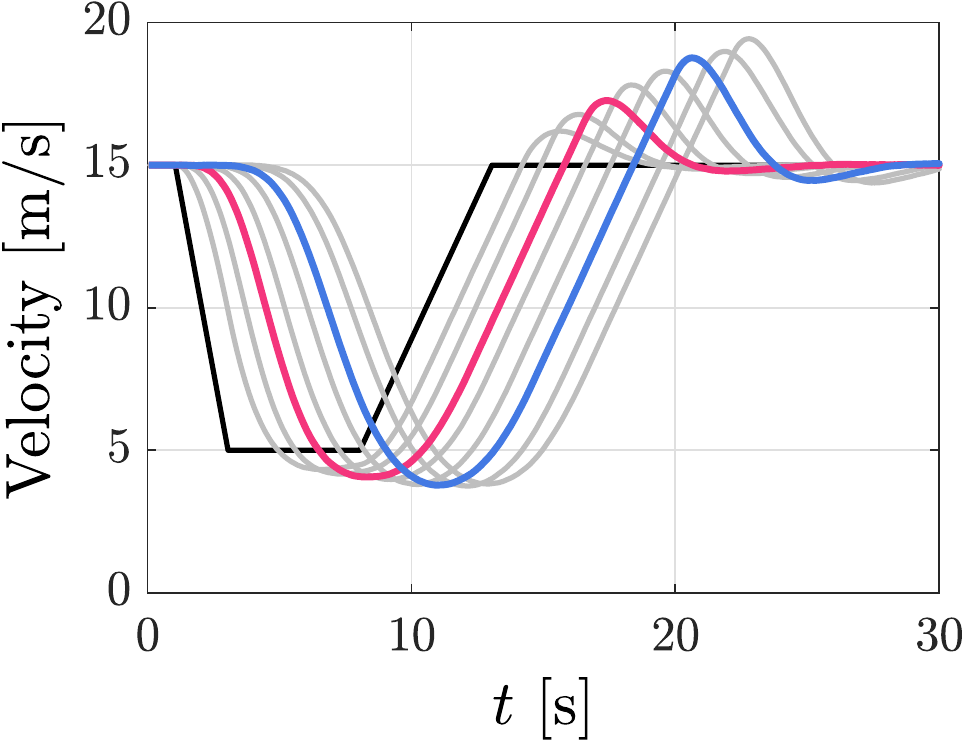}
	\label{Fig:BrakePerturbation_HDVs_Velocity}}
	\subfigure[]
	{\includegraphics[scale=0.42]{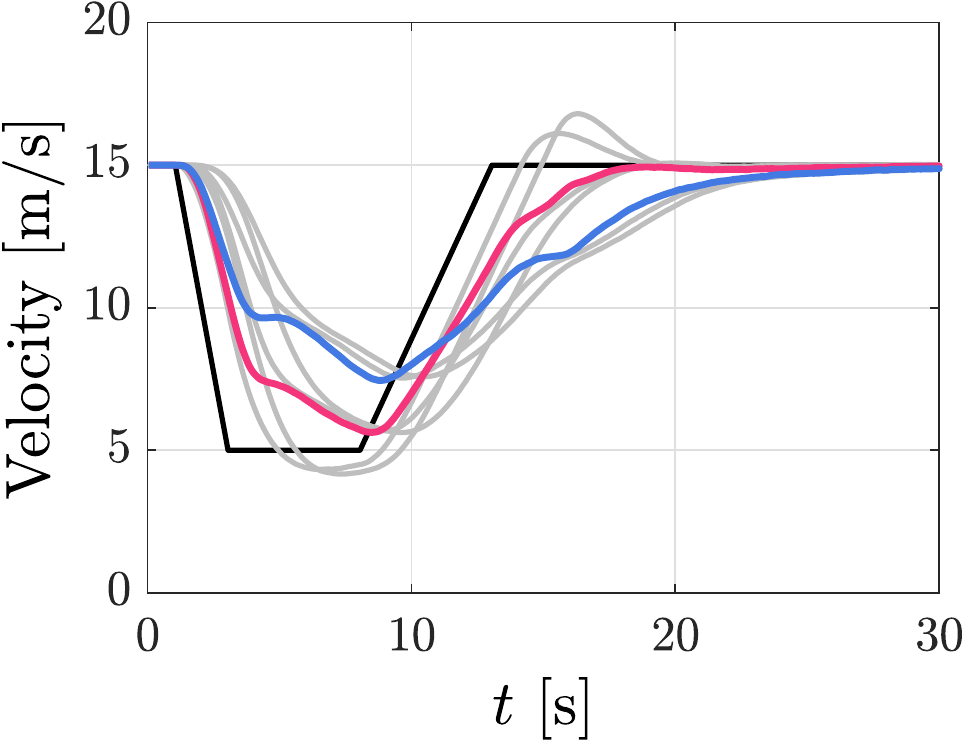}
	\label{Fig:BrakePerturbation_DeePC_Velocity}}
	\subfigure[]
	{\includegraphics[scale=0.42]{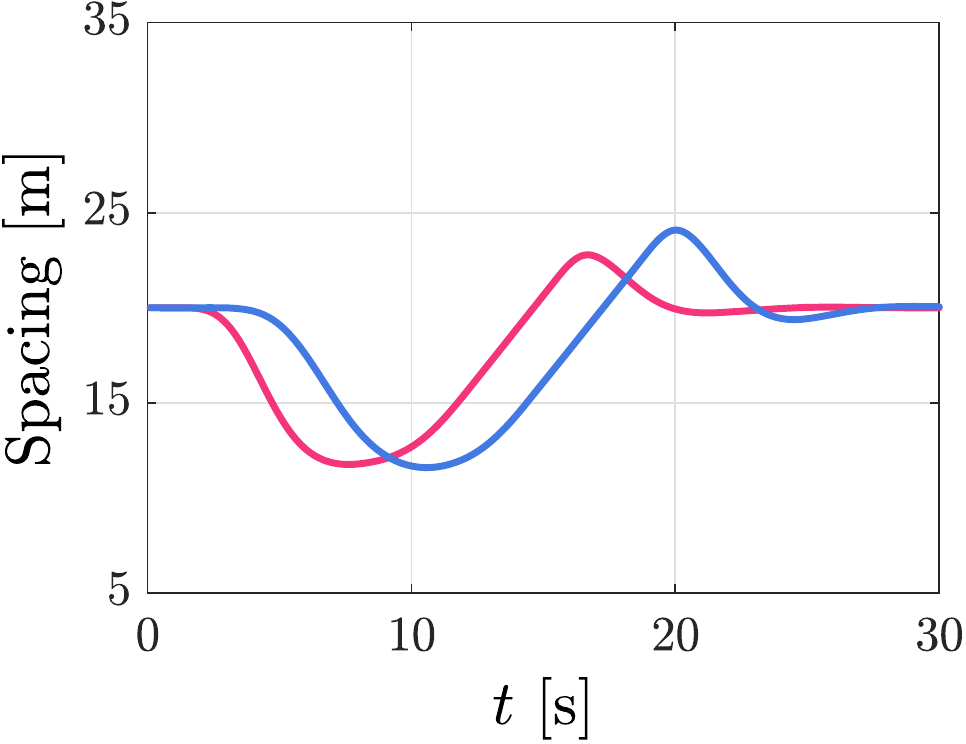}
	\label{Fig:BrakePerturbation_HDVs_Spacing}}
	\subfigure[]
	{\includegraphics[scale=0.42]{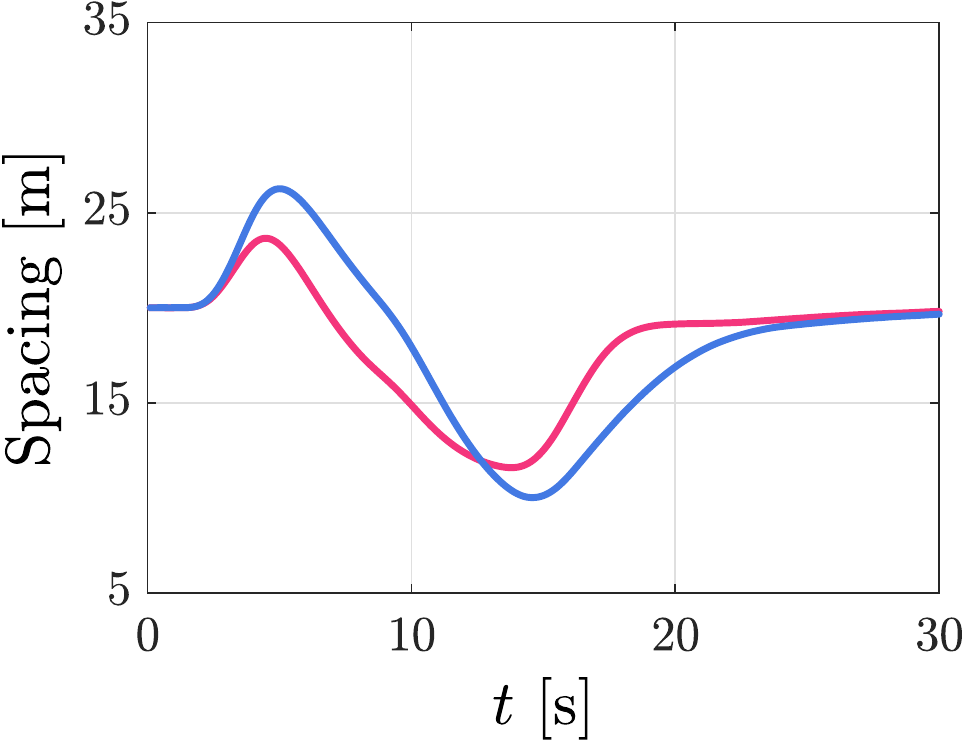}
	\label{Fig:BrakePerturbation_DeePC_Spacing}}
	\subfigure[]
	{\includegraphics[scale=0.42]{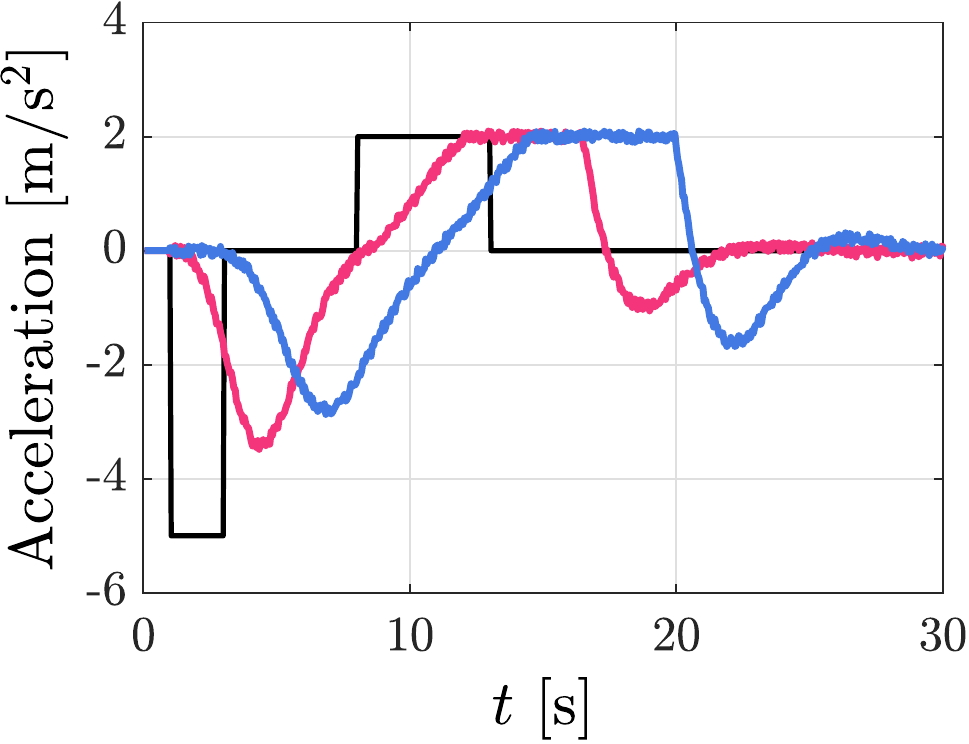}
	\label{Fig:BrakePerturbation_HDVs_Acceleration}}
	\subfigure[]
	{\includegraphics[scale=0.42]{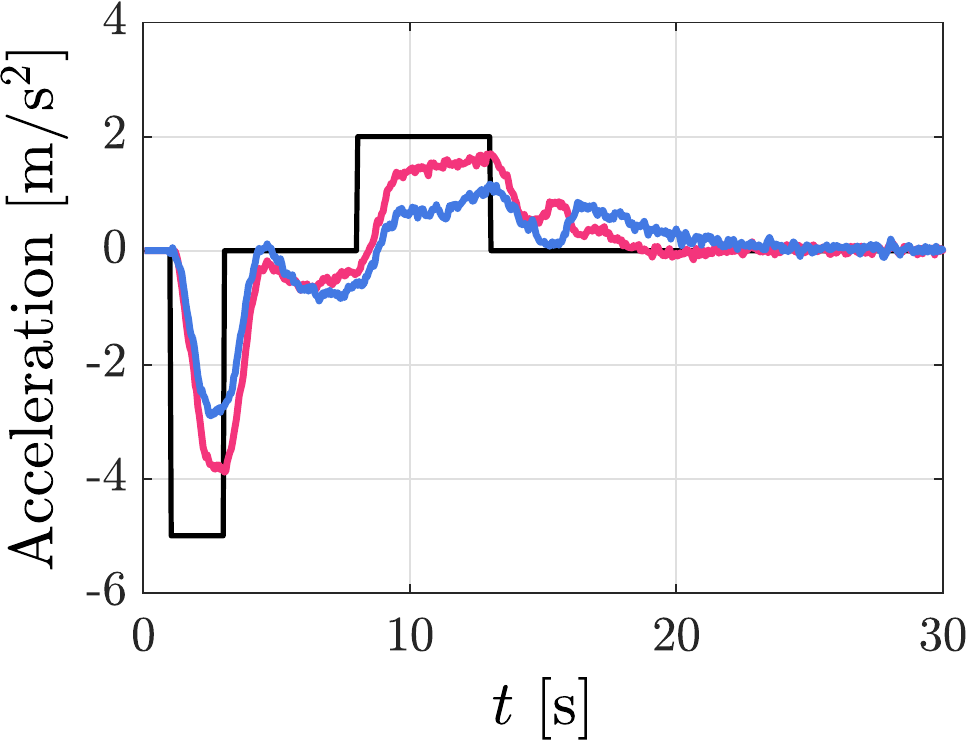}
	\label{Fig:BrakePerturbation_DeePC_Acceleration}}
	\vspace{-2mm}
	\caption{Simulation results in Experiment B. 
	(a)(c)(e) show the velocity, spacing, and acceleration profiles, respectively when all the vehicles are HDVs, while (b)(d)(f) show the corresponding profiles where the two CAVs utilize \method{DeeP-LCC}. In (c)-(f), the profiles of the other HDVs are hided. The color of each profile has the same meaning as that in Fig.~\ref{Fig:NEDCSimulation}.}
	\label{Fig:BrakePerturbation_VelocityProfile}
	\vspace{-5mm}
\end{figure}

The results are shown in Fig.~\ref{Fig:BrakePerturbation_VelocityProfile}. As can be clearly observed, when all the vehicles are HDVs, they have a large velocity fluctuation as a response to the brake perturbation of the head vehicle. By contrast, when two vehicles utilize \method{DeeP-LCC}, they have a quite distinct response pattern from the HDVs. Precisely, the CAVs decelerate immediately when the head vehicle starts to brake, thus leaving a relatively large safe distance from the preceding vehicle (see the time period before $10 \, \mathrm{s}$). Then, the CAVs accelerate slowly when the head vehicle begins to return to the original velocity (see the time period in $9-12\,\mathrm{s}$), while in the case of all the HDVs, they would take a delayed rapid acceleration (see the time period in $12-20\,\mathrm{s}$), which could lead to driving discomfort and collision risk. 
In addition, \textit{$24.96\%$  fuel consumption reduction has been observed} after introducing \method{DeeP-LCC} compared with the case of all HDVs. Our strategy allows the CAVs to eliminate velocity overshoot, improve fuel economy, and constrain the spacing among the safe range, contributing to smoother traffic flow with safety guarantees.

\section{Conclusions}
\label{Sec:6}
In this paper, we have proposed \method{DeeP-LCC} for CAV control in mixed traffic. Our dynamical modeling and controllability/observability analysis guarantee its rationality. In particular, \method{DeeP-LCC} relies directly on the HDVs' driving data rather than a parametric HDV model to design the control input of the CAVs, and it is applicable to nonlinear and non-deterministic traffic flow. Traffic simulations have shown its significant improvement in traffic efficiency and fuel economy, with safety guarantees. Some interesting future directions include incorporating delayed trajectory data caused from communication delay, exploring the scalability of \method{DeeP-LCC} for large-scale mixed traffic, and addressing the problem of time-varying traffic equilibrium.








\bibliographystyle{IEEEtran}
\bibliography{IEEEabrv,mybibfile}

\end{document}